\documentclass[]{article}
\usepackage{booktabs} 
\usepackage{graphicx}
\usepackage{tabularx}
\usepackage{multirow, multicol}
\usepackage{subcaption}
\usepackage{afterpage}
\usepackage{amsmath,amssymb,amsthm}
\usepackage{rotating}
\usepackage{fancyhdr}
\usepackage{tikz}
\usepackage{rotating}
\usepackage{float}
\usepackage{cleveref}
\usepackage[english]{babel}

\newtheorem{theorem}{Theorem}

\newtheorem{definition}{Definition}

\newtheorem{proposition}{Proposition}

\DeclareMathOperator*{\argmax}{arg\,max}

\begin{document}
\title{How to Maximize the Spread of Social Influence: A Survey}

\author{Giuseppe De Nittis and Nicola Gatti \\ \\ Politecnico di Milano \\ Dipartimento di Elettronica, Informazione e Bioingegneria \\ \{giuseppe.denittis, nicola.gatti\}@polimi.it }
\date{}
\maketitle
\ \\
\ \\
\ \\
\begin{abstract}
This survey presents the main results achieved for the influence maximization problem in social networks.
This problem is well studied in the literature and, thanks to its recent applications, some of which currently deployed on the field, it is receiving more and more attention in the scientific community.
The problem can be formulated as follows: given a graph, with each node having a certain probability of \textit{influencing} its neighbors, select a subset of vertices so that the number of nodes in the network that are influenced is maximized.
Starting from this model, we introduce the main theoretical developments and computational results that have been achieved, taking into account different diffusion models describing how the information spreads throughout the network, various ways in which the sources of information could be placed, and how to tackle the problem in the presence of uncertainties affecting the network.
Finally, we present one of the main application that has been developed and deployed exploiting tools and techniques previously discussed.
\end{abstract}

\section{Introduction}
Influencing people is a very ancient issue. All populations in all eras dealt with it.
Going back to the Ancient Greece, there were the \textit{Sophists}, philosophers and teachers, but also incredible speakers. They were so well trained they could take any position and defend any hypothesis.
The Latins had the \textit{Oratori}, like Cicero, involved in politics and very able to move the sentiment and influence people and governors to accept or fight for laws and rights.  
Then, in the Medieval Age, preachers were so inspiring they could move entire masses of people to fight in foreign lands, as happened during the Crusades.
Nowadays, the problem of influencing people has still a political aspect, but it has also acquired a very important commercial feature, e.g., companies investing lots of money in advertising to sell more products and become more and more popular and wealthy.
From a scientific (and computer science) point of view, the Influence Maximization Problem (IMP) has been formalized for the first time in 2003~\cite{kempe2003maximizing}.

\subsection{From Mass to Direct to Viral Market}
Marketing has always been a fertile soil for applications of theories coming from \emph{Knowledge Discovery and Data mining} (KDD)~\cite{fayyad1996data}. 
Instead of focusing on mass marketing, where products are proposed to all the possible consumers without taking into account any further information, like her preferences, KDD studied problems related to \textit{direct marketing}, i.e., among all the possible potential customers, first select the ones that seems more willing to buy the product that is being promoted~\cite{ling1998data}.
Behavioral models have been built to predict the behavior of the customers, e.g., if they will buy some product while not purchasing another one, exploiting information about the client herself and her previous purchases~\cite{domingos2001mining,richardson2002mining}.
However, such an approach has a strong drawback, namely, assuming someone purchases a product only taking into account her own preferences, while people are always influenced by others' opinions, tastes and behaviors.
This kind of marketing, based on word-of-mouth, is called \textit{viral} because it spreads like an epidemic, from person to person~\cite{domingos2005mining,subramani2003knowledge}. 
Viral marketing has two remarkable features. 
First, it is very cost-effective, since the customers propagate the benefits of a product without the direct intervention of a company.
Then, because of of information diffuses according to it, it does not require to build a profile for each customer, but the problem can be reformulated as figuring out who are the most \textit{influential customers}, i.e., people that other people will listen to, and will be driven by in performing their purchases.

While nowadays it may seem obvious, such an approach has been a breakthrough since ignoring these effects and just relying on the preferences of the single user clearly leads to suboptimal solutions. 
Thus, from a practical point of view, the intrinsic value of a customer is not the only one that defines her: even though, if considered \textit{alone}, her value could be smaller than the one of another customer, including the potential value of her network and the possibility of spreading the word and influencing others, may significantly increase her actual value for some marketing campaign.

Unfortunately, quantifying the value of the neighbors of a customer is a hard task since it does not depend only on the person but, potentially, on the entire network. 
This is why all the studies in this research field aim at finding interesting individuals in social networks by exploring and exploiting both the relationships each person has and the topology of the network.

\subsection{Structure of the Paper}
In this work, we focus on the Influence Maximization Problem in social networks. 
The survey about models and algorithms for social influence analysis briefly discusses IMP~\cite[Section 3]{sun2011survey}, presenting some diffusion models and few applications.	
Our paper focuses entirely on IMP, exploring the most important diffusion models that have been proposed, the algorithms developed to solve the problem faster and faster, and the refinements introduced to design models closer and closer to reality. 
The rest of the paper is organized as follows.

\begin{itemize}
	\item \Cref{sec:problem_definition} presents the problem of influence maximization in social networks. First, in~\Cref{sec:influence_max_prob}, we formally present the basic version of the problem of influence maximization. Then, we focus on different extensions that have been developed. Specifically,~\Cref{sec:diffusion_models} reports the main diffusion models that have been proposed,~\Cref{sec:seeds_placement} analyzes different ways of placing the diffusion seeds, and finally, in~\Cref{sec:network}, we present the main uncertainties that may affect the structure of the network.
	\item \Cref{sec:results} provides the most important results achieved for the different scenarios presented in the previous section.
	\item \Cref{sec:application} presents one of the current deployed most interesting applications  developed by implementing methods and techniques presented in the paper. Specifically, the problem is preventing the diffusion of HIV in youth by identifying peer leaders in communities so that, after being trained, they could influence the peers and spread these positive information on how to prevent the contraction of such a disease.
	\item Finally,~\Cref{sec:future} concludes the work and proposes some future directions that could be undertaken.
	\item \Cref{appendix:notation} reports the main symbols adopted throughout the work.
\end{itemize}

\section{Maximizing the Spread of Social Influence}\label{sec:problem_definition}
We introduce the problem of influence maximization as it was formulated for the first time in 2003 in~\cite[Section ~2.1]{kempe2003maximizing}. 
In the rest of the section, we present the main contributions that have been designed building up on this model, grouping them according to the direction they developed.

\subsection{Influence Maximization Problem}\label{sec:influence_max_prob}
Here, we introduce the main elements characterizing a social network and the problem of maximizing the spread of the information.

The goal of the influence maximization problem is to maximize the number of individuals that are reached by some information. For example, we can consider a social network, where the different users are connected by friendship relationships. Here, one may be interested to study how the adoption of some product is influenced by the fact that some friends have already adopted it. 
Another fundamental element is the way in which the influence diffuse in the network, moving from one node to another one. These ways are called \textit{diffusion models}, and we will present several of them that have been developed to be more and more realistic~\cite{kempe2003maximizing,ma2008mining,xie2015dynadiffuse}.

People in the network are usually divided into two main categories: active users, who have been reached by the information, and inactive users, who are not aware of the information yet, but there is a positive probability they may be influenced.
Active users can influence inactive users, changing their status and activating them. One possible way for a node to be activated is because one of her neighbors has been activated and thus, in the next time window, it will try to influence her.
Once a node has been influenced, it will remain active until the end of the process and, unless differently specified, each node tries to influence its neighbors \textit{only} in the time instant after its activation.

Formally, the social network is modeled as a directed graph $G=(V,E)$, where each node $v \in V$ represent an individual in the network and each edge represents the connections between agents. We denote $|V| = n$ and $|E| = m$.
Each edge is characterized by some probability: $p(e): E \rightarrow [0,1]$,~\footnote{Throughout the paper, we will indifferently adopt either $p(e)$ or $p(v,v')$ to denote the influence probability of $v$ w.r.t. $v'$, for each $e=(v,v') \in E$.} indicating the propagation probability along edge $e$. 
In other words, $p(e)$ represents the strength of the relationship involving the agents, e.g., the strength of a friendship.
There is a set $S$ of $k < |V|$ nodes, called \textit{seeds} from which the diffusion starts. The term \textit{seed} has been adopted since they are the starting point from which the information is generated.
Finally, according to some diffusion model $M$, an influence function $\sigma(S): V \rightarrow \mathbb{N}$ is defined. $\sigma(\cdot)$ computes the expected number of nodes that will be reached by the information if the seeds are nodes in $S$.
Our goal is finding the optimal set of seeds $S^* = \argmax_S \sigma(S)$.

The decision version of the problem can be written as follows.

\begin{definition}[Influence Maximization Problem (IMP)] 
	The decision version of Influence Maximization Problem is defined as follows.
	\begin{itemize}
		\item INSTANCE: a graph $G=(V,E)$, influence probability $p(e) \in [0,1]$ for each $e \in E$, a diffusion model $M$, an integer $k$.
		\item QUESTION: is there a subset $S \subset V$, with $|S| \leq k$, such that, placing the seeds in $S$, according to $M$, all the nodes in the network are influenced?
	\end{itemize}
\end{definition}

Next, we present the main features that have been developed to enhance the basic model just described. 
More precisely,~\Cref{sec:diffusion_models} introduces the different diffusion models that have been adopted, starting from the most common ones, e.g., Independent Cascade model, generalizing them, and introducing more realistic ones.
Then, in~\Cref{sec:seeds_placement}, we discuss the main methods that have been proposed to place the different seeds in the network.
Finally,~\Cref{sec:network} presents variations that have been proposed on the structure of the graph itself, e.g., how to redefine the model if the network itself is unknown.

\subsection{Diffusion Models}\label{sec:diffusion_models}
A fundamental component of IMP is the diffusion model, i.e., the way in which information diffuses throughout the network. In this section, we introduce the most adopted ones, presented in~\cite{chen2011influence,kempe2003maximizing,ma2008mining,xie2015dynadiffuse}.

\subsubsection{Cascade Models}
The influence process of the Independent Cascade (IC) diffusion model works according to a discrete representation of the passing of time. Let us suppose that the set of seeds $S$ has been defined.
At time instant $t=0$, nodes in $S$ are activated. Then, at $t=1$, each node $s \in S$ may activate its neighbors $v \in V \setminus S$ with probability $p(s,v)$, which depends on the strength of their connection.
In general, at time instant $t$ there will be $S_t$ nodes active. At $t+1$, each node in $S_t$ will have the chance to influence its neighbors with probability equal to the weight of the edge connecting them, thus independent by the activation history.
If $s \in S$ succeeds, then its neighbor is added to the set of active nodes $S$. However, if $s$ should fail, no other attempts at influencing its neighbors are possible.
The diffusion process ends when no new nodes are added to $S$, i.e., no other activations are possible.
Notice that, because of the single opportunity each node has to influence its neighbors and the fact that such an influence depends on $p(\cdot)$, the process may end even though not all the nodes in the network may have been influenced.

This model can be generalized: we allow the probability that $s$ succeeds in activating one of its neighbors $v'$ to depend also on the neighbors of $v'$ that have already tried spreading the influence.
Formally, we can define an incremental function $p_v(s,V)$, with $V$ being a subset of $v$' neighbors and $s \notin V$.
The diffusion process is the following. When $s$ tries to activate $v'$, it succeeds with probability $p_v(s,V)$, where $V$ is the set of neighbors that have already tried and failed to activate $v$. 
Of course, we consider only cascade models defined by incremental functions that are \emph{order-independent}, i.e., if $s_1,\ldots,s_k$ try to activate $v$, then the probability that $v$ is activated is independent by the order with which such attempts have been performed by $s_1,\ldots,s_k$.
This model is known as \emph{Generalized Cascade} model.

Given the intractability of such models in most of the situations (as we will see in~\Cref{sec:results}), we can add a condition to the cascade models to make them tractable, defining the \emph{Decreasing Cascade} model, in which the probability that some seed $s$ influence some vertex $v$ is \textit{non-increasing} as a function of the sets of nodes that have previously tried to influence $v$.
Formally, this means that $p_v(s,V) \geq p_v(s,W)$, for every $V \subseteq W$. This is known as the \textit{diminishing influence condition}.

There is a further extension of IC, namely IC-N, i.e., Independent Cascade with Negative states, which has been introduced for the first time in~\cite{chen2011influence}.
Here, the nodes may have three states: positive, neutral and negative.
At the beginning, each node $v \notin S$ is neutral. As for IC, a node can be activated at any turn $t$, but in this case can be influenced either positively or negatively.
Thus, at $t=0$, each seed $s \in S$ is positively activated with some probability $q$ and negatively with probability $1-q$, independently on the others.
At $t=t'$, for each neutral node $v$, a permutation of all its neighbors $v'$ activated at time $t'-1$ is computed. Then, according to such permutation, the nodes try to positively or negatively influence $v$ according to their status, with probability $w_{v',v}$, i.e., the influence that $v'$ has on $v$.
As for IC, the process ends when a fixed point is reached, i.e., there are no more activation in the current time $t$.

\subsubsection{Threshold Models}
According to the Linear Threshold (LT) diffusion model, each vertex $v$ has a threshold $\theta_v \in [0,1]$ representing the total weight its neighbors should achieve in order for $v$ to be activated.
The diffusion process develops in a discretized deterministic fashion.
At the beginning, the seeds $s \in S$ are the only active nodes. Then, at step $t$, all the inactive nodes for which the following condition hold:

\begin{equation}
	\sum_{v' \rightarrow v, v' active} p(v',v) \geq \theta_v
\end{equation}

are activated and added to $S$, with $v'$ being neighbors of $v$. Active nodes remain active.
The process stops when no more nodes can be activated. 
From a qualitative point of view, the threshold may represent the tendencies of the different people to be convinced of something, e.g., adopting a particular thought about a political issue or embracing a new technology.

This linear model can be generalized, extending the assumption that each individual gathers influence only linearly.
Specifically, we could associate to each node $v$ an \emph{arbitrary monotonic function} of the set of its already active neighbors. 
Consequently, each node has a monotonic threshold function, $f_v$, such that, when $f_v(v') \geq \theta_v$, where $v'$ are the neighbors of $v$. If this condition holds, then $v$ results activated and it is added to $S$.
The diffusion process is the same as before.
This model is known as \emph{Generalized Threshold} model.

\subsubsection{Triggering Model}\label{sec:triggering_model}
In the triggering diffusion model, each node $v$ independently chooses a random \textit{triggering set} $T_v$ according to some distribution over subsets of its incoming neighbors. 
At the beginning, a set $S$ of seeds is activated.
Then, an inactive node $v$ becomes active in step $t$ if it has a neighbor in its chosen triggering set
$T_v$ that is active in step $t-1$.

Compared to the Threshold model, $v$'s threshold $\theta_v$ has been replaced by a latent subset $T_v$ of
neighbors whose behavior actually affects $v$. 
A nice way of looking at the model is by distinguishing the edges between \textit{live} and \textit{blocked}, depending on whether they belong to the triggering set $T_v$ or not.
Specifically, if $v' \in T_v$, i.e., the triggering set of node $v$ , then the edge $(v',v)$ is dubbed \textit{live}, otherwise it is \textit{blocked}. 
Since our goal is maximizing the number of active nodes \textit{at the end} of the diffusion process, then we are saying that $S_n$, i.e., the set of active nodes at time $t = n$, is the set of nodes $v$ such that $v$ is reachable from $S$ via a path consisting entirely of live edges.

\subsubsection{Heat Diffusion Model}
This interesting diffusion model has been proposed for the first time in~\cite{ma2008mining}. The rationale behind is exploiting the physical phenomenon of how the heat diffuses from a point with a higher temperature to a point with a lower temperature. 
The same principle can be applied to social influence and the spreading of, for example, innovations: in this scenarios, the \textit{innovators} are the heat sources and have a high temperature, so that their ideas diffuse to other people throughout the network, as heat diffuses form point to point in some medium.
Formally, the Heat Diffusion Model (HDM) can be described as follows.~\footnote{For the sake of presentation, here we will introduce only the heat diffusion model for undirected graphs. See~\cite[Sections~3.2--3.3]{ma2008mining}, for further details on directed graphs with and without prior knowledge of the diffusion probability.}

Let $\theta_v$ be the activation threshold of node $v$, and let us denote with $h_v(t)$ the heat at node $v$ at time instant $t$. Let us assume that at time $t=0$, the heat distribution starts, and, at $t=t'$, each node $v$ receives some amount of heat from each of its neighbor $v'$ during the period $\Delta t$.
It is reasonable to assume that the heat received by $v$ will be proportional to the heat difference between $v$ and $v'$ and time $\Delta t$.
Thus, we can formulate this as:

\begin{equation*}
	\frac{h_v(t'+\Delta t) - h_v(t')}{\Delta t} = \alpha \cdot \sum_{\forall v'|(v,v') \in E} (h_{v'}(t')-h_v(t')) = \alpha \boldsymbol{H}h(t),
\end{equation*}

where $\alpha$ is the heat diffusion coefficient and $H$ is defined as:

\begin{equation*}
	\boldsymbol{H} = \begin{cases} 1, & \mbox{if }(v,v') \in E\\
		-d(v), &\mbox{if }(v = v')\\
		0, & \mbox{otherwise.}
	\end{cases}
\end{equation*}

In the above formula, $d(v)$ is the degree of node $v$.~\footnote{Since we defined the Heat Diffusion Model on undirected graphs, the outgoing degree of some node $v$ is equal to its ingoing degree.} 
If $\Delta t \rightarrow 0$, then we obtain:

\begin{equation*}
	\frac{d\,}{dt\,}h(t) = \alpha \boldsymbol{H}h(t).
\end{equation*}

Solving this equation, we finally get:

\begin{equation*}
	h(t) = e^{\alpha t \boldsymbol{H}}h(0),
\end{equation*}

where $e^{\alpha t \boldsymbol{H}}$ is called \textit{diffusion kernel} since the heat diffusion process continues infinitely from the initial diffusion at $t = 0$. 
If the amount of heat received by $v$ is at least $\theta_v$, then the node is activated.

\subsubsection{Dynamic Diffusion Model}
This model has been proposed to enrich the diffusion model by making it dynamic~\cite{xie2015dynadiffuse}.
The novelty here is that they adopt the exponential distribution to model the propagation rate since different edges may have different ones, and they may change over time.
Formally, for the Dynamic Diffusion model (DynaDiffuse), let us consider two adjacent nodes, say $v, v'$. Then, the influence diffusion probability is $1-e^{rt}$, where $t$ is the timespan after $v'$ has been influenced and $r$ is the \textit{propagation rate}.
Differently from the other models, here each node tries repeatedly to influence its neighbors. 
With these elements, we can infer the diffusion network, and a Continuous Time Markov Chain can be built, provided an initial set of vertices $I$.

\begin{definition}[Continuous Time Markov Chain (CTMC)]
	A CTMC is a tuple $(\zeta,\varsigma_0,R,L)$, where $\zeta$ is a set of states and $\varsigma_0$ the initial one. $R: \zeta \times \zeta \rightarrow R_{\geq 0}$ is a transition rate matrix assigning rates to pairs of states, with $r \in R$ used as the rate parameter for the exponential distribution $f(r) = re^{-rt}$, for $t>0$. For a fixed finite set $P$ of action labels, $L: R \rightarrow 2^P$ assigns such labels to every transition with $r>0$.
\end{definition}

To represent the diffusion according to DynaDiffuse, we build a CTMC. 
Specifically, each state $\varsigma \in \zeta$ has $|V|$ dimensions and an associated boolean value saying whether or not the node has been activated. The construction is as follows.

\begin{itemize}
	\item Create an initial state $\varsigma_0$ for $I$, the initial set.
	\item Then, iteratively create all other reachable states in $\zeta$ and the corresponding transitions in $R$. If some states are already in $\zeta$, just add the transitions.
	\item Repeat the previous point until there is nothing more to add.
	\item For a transition from $\varsigma_i$ to $\varsigma_j$, the transition state is computed as $\sum_{e_i} r_i$, where $r_i$ is the propagation rate of each edge that can influence the new node.
\end{itemize}

The last step is introducing additional variables to model the dynamic properties of the networks.
Specifically, local variables $f_1,\ldots,f_m$ are introduced to model the degree of effects on the diffusion network by a positive function $\Phi(f_1,\ldots,f_m)$.
In a CTMC, transitions divided into two groups: \textit{internal}, which define how these dynamic factors evolve stochastically, and \textit{external}, which are associated with action labels that will be used for synchronization among CTMCs.

CTMC evolves in time according to action labels. Moreover, transitions with different labels in different CTMC may occur independently. 
With this formalism, we can model the most common dynamic characteristics. Thus, we get a labeled CTMC $M_0$ for an inferred diffusion network and a series of CTMCs $D = \{d_1,\ldots,d_n\}$ for the dynamics. Finally, we denote with $M_1 \| M_2$ the parallel composition of two labeled CTMCs.~\footnote{Please, refer to~\cite[Section 4]{xie2015dynadiffuse} for more details on CTMCs.}
Now we are ready to define DynaDiffuse.

\begin{definition}[Dynamic Diffusion (DynaDiffuse) model]
	Given a CTMC $M_0$ for a network $G=(V,E)$, a set $A$ of propagation rates $\alpha_{v,v'}$ for each $(v,v') \in E$, an initial set of vertices $I$ and a set of dynamic characteristics $D = \{d_1,\ldots,d_n\}$, where each $d_i \in D$ is a CTMC model whose transition matrix represents a positive function $\Phi(f_1,\ldots,f_m)$, DynaDiffuse is a synchronous CTMC model defined as $M = M_0 \| D = M_0 \| d_1 \| \cdots \| d_n$.
\end{definition}

\subsection{Seeds Placement}\label{sec:seeds_placement}
In this section, we introduce the main models that have been proposed to extend the possible ways of placing the diffusion seeds in the network.
Specifically, we first present a model in which only \textit{jump} and \textit{crawl} operations are allowed, and then we will deal with the \textit{adaptive seeding} problem.

\subsubsection{Jumping and Crawling}
Considering the ways in which individuals can be reached by other people on social networks, we analyze scenarios where we are allowed to perform only two operations, which are the main ones that may be undertaken in social networks~\cite{brautbar2010local}.
The first action is \emph{crawling}: if a user has been discovered or added to our friends, we are usually provided also the links to her neighbors, e.g., friends/followers.
The second action we can perform is \emph{jumping}: in general, we are given the possibility of searching some other node which may be far from the discovered ones, e.g., by means of a search bar allowing us to find new users. While in real-world applications such a search is not random, i.e., there are different probabilities for us searching new users, for simplicity we assume that, once we execute the jump, each node has the same probability of being selected. 
In such settings, the goal is still finding the most influential nodes of the network, however such a search is performed w.r.t. two main parameters.
First, the degree $d(v)$ of the nodes is considered, with the aim of finding nodes with the highest degree.
Formally, let $d^*$ be the maximum degree of the different nodes. Then, given $0 < \beta < 1$, we want to find a vertex $v$ such that: $d^* \leq \textnormal{d}(v)n^{1-\beta}$.

Beside considering the maximum degree $d^*$, the authors also investigated the clustering coefficient, aiming at finding nodes with the highest one.

\begin{definition}[Clustering coefficient]
	Let $v$ be a vertex with degree $d$ of a graph $G$. Then, the Clustering Coefficient (CC) of $v$ is:
	
	\begin{equation*}
		CC(v)= n_{\Delta} \bigg/ \frac{d}{2}
	\end{equation*}
	
	where $n_{\Delta}$ is the number of cliques of dimension $3$ of $G$.
\end{definition}

Informally, given some vertex, CC measures how densely connected its neighbors are, i.e., their tendency to form a clique.

Finding the node with the highest CC may not be very informative since it may happen that the vertex with the highest CC has only few neighbors, thus not being interesting to find potential influential nodes in the network. This is why we look for nodes that have both a high CC and degree. Specifically, given some degree lower bound $d^-$, we want to find a vertex of degree not smaller than $d^-$ whose CC approximates the maximum CC among all vertices of degree $d$ or bigger.

\begin{definition}[Approximation of CC]
	Given a graph with $n$ vertices and some degree value $d$, let $v^*$ be the vertex with the highest CC among vertices of degree $d$ or more. Then, $v$ is an $(\alpha,d,\epsilon)$-approximation to the maximum CC if $d(v) \geq \alpha d$ and $CC(v^*) \leq CC(v)+\epsilon$, with $0 < \alpha \leq 1$, $0 < \epsilon < 1$.
\end{definition}

\subsubsection{Adaptive Seeding}
Let us consider another interesting way of placing seeds, based on the so-called \textit{friendship paradox}, first discovered by~\cite{feld1991your}.

\begin{proposition}
	In any network, the expected degree of a node is bounded from above by the expected degree of a neighbor.
\end{proposition}

This result is saying that while a great majority of nodes will be ineffective in terms of maximizing the diffusion of the influence, selecting their neighbors could be more convenient.
Thus, instead of investing the entire budget to immediately select $k$ possible seeds, it would be better to adopt a two-stage seeding approach~\cite{seeman2013adaptive}.
First, some budget is used to select a starting set of seeds $S \subseteq V$, making their neighbors accessible. Then, the remaining budget is used to select a set of other seeds from such a larger pool.
Of course, the best policy is the one that, while minimizing the first $k$ chosen nodes, also makes the other most influential nodes of the network accessible. 
Observe that despite the the first stage being an actual problem of influence maximization, the second stage is not, since $\sigma(\cdot)$ models a word-of-mouth process that happens without any incentive.

Before proceeding to the next section, we want to point out a final remark about the \textit{multi-stage} nature of this approach. 
Given such a two-stage approach, one could wonder whether a multi-stage approach could be beneficial, e.g., performing these selection in three or more stages. 
However, the authors empirically show that increasing the number of stages provides only a very marginal contribution to the quality of the solution w.r.t. the time required for computing such a solution. Moreover, they conjecture that optimizing a multi-stage process is computationally hard.

\subsection{Network Uncertainties}\label{sec:network}
In this section, we present scenarios involving uncertainties of the social network.
Specifically, first we will focus on maximizing the spread of social influence in unknown graphs~\cite{mihara2015influence}. 
Then, we will study IMP when there is uncertainty on the probability function $p(\cdot)$. On one side, we will present a \textit{robust} approach~\cite{chen2016robust}, on the other, we will discuss how to deal with this problem in an \textit{online} fashion data are available only as the diffusion process develops, adopting a \textit{multi-armed bandit} approach~\cite{vaswani2015influence}.

\subsubsection{Unknown Network}
The novel feature we present here is tackling IMP without knowing the structure of the social network~\cite{mihara2015influence}.
Most of existing algorithms assume that the entire topology of the network is given and so the goals are only scalability for the exact approaches and the quality of the solution for the approximated ones. 
However, in these settings an algorithm to compute the best placements for the seeds should also explore the network in an intelligent way in order to reconstruct the presence of the various edges.

We are given a graph $G=(V,E)$ modeling the network, but $E$ is unknown, and only a small number of probes is allowed to obtain hints about the topology.
There are $R$ rounds and in each round $\eta$ nodes can be probed, i.e., when node $v$ is probed, it will return the list of its neighbors. Then, as is customary, $k$ nodes are selected as seeds and, at the end of the influence process, the number of active nodes is computed.
The goal is to maximize the expected number of influenced nodes at the $R$-th turn, considering that, at each round, both the probing and the influence spreading results of the previous rounds are available. 
Such an assumption is relaxed in~\cite{wilder2018maximizing}.

\subsubsection{Uncertainty on the Influence}
Another type of uncertainty that has been studied concerns the weights of the edges, i.e., the influence power that individuals have on each other. 
In other words, we know some edges exist, but we do not know how intense the influence spreads. We present two main approaches to two slightly different problems.
First, we present the online version of such a problem\cite{vaswani2015influence}, pointing out the new features w.r.t. the previous models and which techniques should be employed to solve it. 
Then, we discuss a \textit{robust} approach to the problem~\cite{chen2016robust}, with the goal of maximizing the worst-case ratio between the expected number of influenced nodes reached by our choice for the placement of the seeds w.r.t. the optimal choice.

\paragraph{Online Influence Maximization}
Up to now, we considered models that have complete information on the structure of network. 
However, in reality, it is very difficult, if not impossible, being able to access this kind of information or being accurate in quantifying it. 
To avoid this assumption, in~\cite{vaswani2015influence}, the authors adopt a novel approach based on a combinatorial multi-armed bandit paradigm~\cite{chen2013combinatorial}, estimating the probabilities on the edges round by round, selecting a different seed set each time.
The goal is to minimize the accumulated regret incurred by choosing suboptimal seed
sets over multiple rounds. 
The authors consider two possible feedbacks: \textit{edge-level feedback}, which allows to observe how the influence spread among the different edges throughout the network, and \textit{node-level feedback}, that assumes only to know whether an individual has become active or not, without knowing who influenced it.
We now report some information about Combinatorial Multi-armed Bandit (CMAB) that will be useful in the following.

In CMAB framework there are $m$ arms, each associated with a random variable, denoted as $X_{i,t}$, indicating the reward of triggering arm $i$ on round $t$.
$X_{i,t}$ is defined on $[0,1]$ and is independently and identically distributed according to some unknown distribution, with mean $\mu_i$.
At each round $t \in T$, a set of arms, called \textit{superarm} and denoted by $A$, is played, triggering all the arms in the set.
Moreover, some other arms may be probabilistically triggered.
Let $p^i_A$ be the triggering probability of arm $i$ if the superarm A is played (of course, if $i \in A, p^i_A = 1$). 
The reward obtained at each round $t$ is a function of the rewards of the arms that have been triggered in $t$.
Each time an arm $i$ is triggered, we observe the rewards and update its mean estimate $\hat{\mu_i}$
The superarm that is expected to give the highest reward is selected in each round by an oracle $O$, which takes as input the current mean estimates and outputs an appropriate superarm A.

\paragraph{A Robust Approach}
The main feature characterizing such new model is that the influence probabilities of the edges belong to intervals rather than being precise values.
A classical approach in the literature would be applying learning methods to extract the edge probabilities~\cite{goyal2010learning,netrapalli2012learning,rodriguez2011uncovering,saito2008prediction,tang2009social}. However, due to data limitation, no learning method could extract the exact values of the edges probabilities. Thus, the best solution is looking for estimates of such weights together with some confidence intervals. Despite this other approach, the uncertainty affecting the estimates may significantly influence the subsequent influence problem in a negative way.
The authors adopt a learning method considering as input for the maximization problem intervals in which the true weight of the edge may lie~\cite{chen2016robust}.

To generate a random live-edge graph, we say that some edge $e$ is live if flipping a biased random
coin with probability $p(e)$ returns success, otherwise $e$ is called blocked (with probability $1-p(e)$).~\footnote{As the careful reader may notice, this is exactly the same terminology we employed to describe the Triggering Diffusion model in~\ref{sec:triggering_model}.}
We call $L$ the subgraph consisting of all the vertices in $V$ and the live edges in $E$.
Let $S$ be the set of seeds and $R_L(S) \subseteq V$ the reachable nodes from $S$ in $L$, i.e., the set of nodes that can be reached only walking live edges.
Finally, $\theta=(p(e))_{e \in E}$ denotes the probabilities on all the edges. 
For these instances, it is very difficult to obtain true values for the weights of the edges. However, it is possible to get the statistically significant neighborhood $[l_e,r_e]$, i.e., the confidence interval in which $p(e)$ lies. Of course, we cannot tackle the IMP only knowing such intervals since this approach could lead to a huge loss. This is why we resort to a \textit{robust} approach.

Formally, let $\Theta=\times_{e \in E} [l_e,r_e]$ be the parameter space of $G$, with  $\theta=(p(e))_{e \in E}$ being its latent parameter vector. 

\begin{definition}[Robustness ratio]
	For a seed set $S \subseteq V$, with $|S| = k$, the robust ratio under parameter space $\Theta$ is:
	\begin{equation*}
		\rho(\Theta,S) = \min_{\theta \in \Theta} \frac{\sigma_{\theta}(S)}{\sigma_{\theta}(S^*_{\theta})}
	\end{equation*}
	
	where $\sigma_{\theta}(\cdot)$ is the number of expected influenced nodes at the end of the spreading process and $S^*_{\theta}$ denotes the optimal solution of size $k$ and the probability of every edge is $\theta$.
\end{definition}

We observe that given $\Theta$ and a solution $S$, $\rho$ embodies the worst-case ratio w.r.t. the number of influenced nodes between $S$ and the optimal placement of seeds in $G$. 
Now we are ready to formulate the Robust Influence Maximization Problem.

\begin{definition}[Robust Influence Maximization Problem (RIMP)]
	The problem of Robust Influence Maximization is defined as:
	\begin{itemize}
		\item INSTANCE: a graph $G=(V,E)$, weights $p(e) \in [0,1]$ for each $e \in E$, a parameter space $\Theta=\times_{e \in E} [l_e,r_e]$, and an integer $k$.
		\item QUESTION: find a subset $S \subset V$, with $|S|=k$, such that the robust ratio $\rho$ is maximized, i.e.,
		
		\begin{equation*}
			S^*_{\theta} = \argmax_{S \subseteq V, |S|=k} \rho(\Theta,S) = \argmax_{S \subseteq V, |S|=k} \min_{\theta \in \Theta} \frac{\sigma_{\theta}(S)}{\sigma_{\theta}(S^*_{\theta})}
		\end{equation*}
	\end{itemize}
\end{definition}

Notice that when there is no uncertainty in the intervals, i.e., $\Theta$ collapses on the true probabilities $\theta$, then RIMP coincides with its exact counterpart, IMP.

We highlight a final remark on the notion of robustness. Beside the meaning we just described, in~\cite{he2016robust} such a term is adopted w.r.t. the guarantees that an algorithm may provide w.r.t. a spectrum of different diffusion models and parameter settings. We will also deepen this aspect in the following section.

\section{Achievements and Results}\label{sec:results}
In this section we present the most important achievements and results developed for IMP and its extensions.
Specifically, in~\Cref{sec:fundamental_results}, we start reporting the fundamental results, and then in~\Cref{sec:scalability} we focus on how this problem has been solved faster and faster for the different diffusion models. 
Next, we discuss all the results concerning the other main features characterizing IMP, namely, the seeds placement and the uncertainties that may affect structure of the network, in~\Cref{sec:results_seeds,sec:results_network}, respectively. To facilitate the reader, each paragraph has been named after the algorithm proposed by the various authors to tackle the different problems.

\subsection{Fundamental Results}\label{sec:fundamental_results}
We start by studying the complexity of IMP, and then we will discuss the first algorithm that has been proposed. 
For the problem to be tractable, two fundamental assumptions are commonly made on the influence function $\sigma(\cdot)$, namely the monotonicity and submodularity.

\begin{definition}[Monotonic function]
	A function $f$ defined over a domain $X$ is monotonic if either of the following conditions is satisfied.
	\begin{itemize}
		\item If, for all $x,y \in X, x \leq y$, then $f(x) \leq f(y)$.
		\item If, for all $x,y \in X, x \leq y$, then $f(x) \geq f(y)$.
	\end{itemize}
\end{definition}

\begin{definition}[Submodular function]
	A function $f$ defined over a domain $X$ is submodular if the following condition is satisfied:
	\begin{equation*}
		f(A \cup \{x\}) - f(A) \geq f(B \cup \{x\}) - f(b)
	\end{equation*}
	
	for all elements $x \in X$ and $A \subseteq B \subseteq X$.
\end{definition}

In other words, given two sets such that the first one is included in the second one, it is more significant adding an element to the first set w.r.t. the second, since one additional unit will give a higher contribution if the set it is added to is \textit{small}.

Notice that the influence function is monotonic by definition, i.e., adding an element to a set cannot cause the influence function to decrease: if we add another placement for a seed, the expected number of influenced nodes cannot decrease, it will stay constant in the \textit{worst} case. Moreover, the following results hold~\cite{kempe2003maximizing}.

\begin{theorem}
	For any arbitrary instance of the Independent Cascade model, the influence function is submodular.
\end{theorem}

\begin{theorem}
	For any arbitrary instance of the Linear Threshold model, the influence function is submodular.
\end{theorem}

Now, we are ready to state two important theorems.

\begin{theorem}
	IMP is $\mathsf{NP}$-hard for the Independent Cascade model.
\end{theorem}

\begin{theorem}
	IMP is $\mathsf{NP}$-hard for the Linear Threshold model.
\end{theorem}

In both cases, the reduction is from the Set Cover problem, which reads as follows.~\footnote{The reader can observe the similarities between SCP and IMP. For more details, please refer to~\cite{kempe2003maximizing}.} 

\begin{definition}[Set Cover problem]
	The decision problem of Set Cover is defined as:
	\begin{itemize}
		\item INSTANCE: a set of elements $U$, a collection of $n$ sets whose union equals the universe, an integer $k$.
		\item QUESTION: is there a subcollection of $S$ whose union equals $U$ such that its cardinality is at most $k$?
	\end{itemize}
\end{definition}

Since solving IMP is $\mathsf{NP}$-hard w.r.t. both diffusion models, the next step is understanding whether such a problem could be at least approximated and, if so, up to which factor. For IC~\footnote{All the hardness and approximability results reported here for IC also holds for IC-N since, also for this model, the influence function is non-negative, monotonic, and submodular.} and LT models, we can exploit the following result~\cite{nemhauser1978analysis}.

\begin{theorem}
	For a non-negative, monotonic, submodular function $f$, let $S$ be a subset of size $k$ obtained by selecting elements one at a time, each time choosing an element that provides the largest marginal  increase in the function value. Let $S^*$ be a set that maximizes the value of $f$ over all $k$-element sets. Then, $f(S) \geq \left(1-\frac{1}{e}\right)f(S^*)$.
\end{theorem}

The above theorem means that $S$ provides a $\left(1-\frac{1}{e}\right)$-approximation. 
Since we know that both in the IC and in the LT models the influence function is non-negative, monotonic and submodular, we have a positive approximation result.

\begin{theorem}
	Let $S$ be the current set of nodes in which seeds are put. Adopting a greedy approach to solve IMP, i.e., adding to $S$ the node with the highest expected number of influenced nodes, guarantees to obtain an approximate solution to within a factor of $1-\frac{1}{e}$.
\end{theorem}

In other words, starting from the empty set, the algorithm repeatedly adds the node that maximizes $\sigma(S \cup \{x\})-\sigma(S)$. The problem is that it is not clear how to evaluate $\sigma(\cdot)$ in polynomial time, and it has actually been proved that such a task is, in general, $\mathsf{\#P}$-complete, both for IC and LT models~\cite{chen2010scalable}. Nevertheless, we can obtain good approximations of $\sigma(\cdot)$ by simulating the random choices and diffusion process a sufficient number of times. Specifically, the Chernoff-Hoeffding bounds~\cite{habib2013probabilistic} imply the following result~\cite{kempe2015maximizing}.

\begin{proposition}
	If the diffusion process, originated from the seeds placed in $S$, is simulated independently at least $\Omega\left(\frac{n^2}{\epsilon^2}\ln(\frac{1}{\delta})\right)$ times, then the average number of activated nodes over these simulations is a $(1 \pm \epsilon)$-approximation to $\sigma(S)$, with probability at least $\-\delta$.
\end{proposition}

Unfortunately, these results do not for every influence function.

\begin{theorem}
	In general, it is $\mathsf{NP}$-hard to approximate IMP to within a factor of $n^{1-\epsilon}$, for any $\epsilon > 0$.
\end{theorem}

The greedy algorithm has been tested on a graph with $10748$ nodes and edges between $53000$ pair of nodes~\cite{kempe2003maximizing} w.r.t. three different models: LT, weighted IC and IC with uniform probability $p$ on the edges.

The algorithm has been compared with the baseline of choosing random nodes and two heuristics, the first based on nodes' degrees and the second on nodes' centrality in the network.
More specifically, the high-degree heuristic chooses node $v$ in order of decreasing order $d(v)$, while the second one selects nodes in order of increasing average distance to other nodes in the network.

When the LT diffusion model is adopted, the greedy algorithm outperforms the high-degree node heuristic by $18\%$ and the central node heuristic by $40\%$. The main reason why both heuristics perform poorly is that they both ignore that most of the central or high degree nodes may be clustered, so that is it a waste of resources putting seeds in all of such nodes.

When the weighted IC model is adopted, all values are $25\%$ than the previous setting, but the trends are the same.

Finally, for the uniform IC mode, $p$ has been set equal to $0.01$. In this case, the network effects in the IC model with very small probabilities are much weaker than in the other models. Several nodes have degrees well exceeding 100, so the probabilities on their incoming edges are even smaller than $1\%$ in the weighted IC model. This suggests that the network effects observed for the LT
and weighted IC models rely heavily on low-degree nodes as multipliers, even though targeting
high-degree nodes is a reasonable heuristic. 

These results lead us to the fact that knowing the dynamics of a network leads to significantly better results than relying only on the structure of the network itself.

\subsection{Solving IMP Faster and Faster}\label{sec:scalability}
In this section, we analyze the main improvements that have been proposed to enhance the computational speed required to solve IMP. 
Despite being very popular, the greedy approach adopted in~\cite{kempe2003maximizing} does not scale up to realistic problems. Thus, several heuristics have been designed through the years to deal with this computational issue.

One of the main reasons for the algorithm in~\cite{kempe2003maximizing} of being slow is the computation of the influence function $\sigma(\cdot)$ since, as proved in~\cite{chen2010scalable} for the Independent Cascade diffusion model, performing such a computation is $\#\mathsf{P}$-hard.
Such an issue has been tackled by estimating the spread using Monte Carlo simulation or by using heuristics~\cite{chen2009efficient,chen2010lt,chen2010scalable,kimura2006tractable}.

We now present the main heuristics according to the diffusion model they have been designed for.

\subsubsection{Independent Cascade model}

\paragraph{SPM/SP1M}
First,~\cite{kimura2006tractable} considers two special cases of the Independent Cascade model and provides approximation algorithms to deal with them.
\begin{itemize}
	\item[1:] Each node is activated only through the shortest paths from the initial active set. Such a model, called \textit{Shortest-Path Model (SPM)}, is a special type of the Independent Cascade model where only the most efficient information spread can occur.
	\item[2:] SPM is generalized to \textit{SP1 Model (SP1M)}, where each node $v$ has the chance to become active only at steps $t = w(S, v)$ and $t = w(S, v) + 1$, i.e., node $v$ cannot be
	activated excluding the paths from $S$ to $v$ whose length are equal to $w(S, v)$ or $w(S, v) + 1$. Here, $w(v,v')$ is the distance between nodes $v$ and $v'$ in the graph.
\end{itemize}

For both models, they propose a more efficient greedy heuristic than the one proposed in~\cite{kempe2003maximizing}, and provide the following result.

\begin{theorem}
	In the SPM and SP1M, for the greedy algorithm, the following result holds:
	\begin{equation*}
		\sigma(B_k) \geq \left(1 - \frac{1}{e}\right)\sigma(A^*_k)
	\end{equation*}
	
	where $B_k$ is the $k$-element set obtained by the greedy algorithm while $A^*_k$is the set that maximizes the value of $\sigma(\cdot)$ over all $k$-element subsets of $V$. 
\end{theorem}

\paragraph{CELF}
The other major issue of~\cite{kempe2003maximizing} is that the algorithm is quadratic in the number of nodes. 
To solve this problem, a first attempt has been made in~\cite{leskovec2007cost}. 
Here the authors study the problem of \textit{outbreak detection}, i.e., selecting nodes in a network to detect the spreading of information as quickly as possible. This is very close to our goal of finding the best placements for seeds spreading the information. 
They propose CELF (Cost-Effective Lazy Forward selection): it is based on a lazy evaluation of the objective function, i.e., if the contribution a node brought to the influence function in the previous iteration of the greedy algorithm was already smaller than the contribution given by the current node, then such a node should not be reevaluated since, by submodularity, its contribution can only be lower.

\paragraph{RanCas/NewGreedyIC}
Next,~\cite{chen2009efficient} improves the greedy approach for the Independent Cascade model, and proposes a heuristic that achieved better results than CELF. Moreveor, the authors also introduce a new degree discount heuristics that improves influence spread.
Specifically, they propose a novel algorithm to simulate the cascade, called \textit{RandomCascade} RanCas($S$). It works as follows: let $S_t'$ be the set of vertices that are activated in the $t$-th round, with $S_0 = S$.
For any edge $(v,v') \in E$ such that $v \in S$ and $v'$ has not been activated yet, $v'$ can be activated by $v$ with probability $p(v,v')$ in the $(t+1)$-th round. 
If $v'$ has $l$ neighbors, then $v' \in S_{t+1}$ with probability $1 - (1-p(v,v'))^l$.
The process continues until $S_{t+1}$ is empty.

Since in RanCas(S), each edge is determined once, and the probability
on either direction is the same, we could determine whether $(v,v')$ is selected for propagation or not, and remove all edges not for propagation from $G$ to obtain a new graph, say $G'$.
Thus, by randomly generating $G'$ for $R$ times, we can select the next best candidate vertex $v$.~\footnote{For more details, please see~\cite[Section 2]{chen2009efficient}.}

In their new algorithm, NewGreedyIC, each random graph is used to estimate the influence spread of all vertices. Comparing NewGreedyIC algorithm with CELF, there is a tradeoff in running time. 
For CELF, the first round is as slow as the original algorithm, but, starting from the second one, each round may only need to explore a small number of vertices and the exploration of each vertex is typically fast since RanCas(S) usually stops after exploring a small portion of the graph. 
Conversely, in every round of NewGreedyIC algorithm, we need to traverse the entire graph $R$ times to generate $R$ random graphs $G'$. 
To combine the merits of both improvements, the authors compose an additional algorithm, called MixGreedyIC, in which in the first round NewGreedyIC is used to select the first seed and compute influence spread estimates for all vertices, and then in later rounds the CELF optimization is adopted to select remaining seeds. This algorithm experimentally shows the best running time.

\paragraph{MIA}
The next contribution has been provided in~\cite{chen2010scalable}, where the authors propose a heuristic algorithm that gains efficiency by restricting computations on the local influence regions of nodes.

The rationale behind such an approach is that they use local arborescence structures of each node to approximate the influence propagation. We recall that, in a directed graph, an arborescence is a tree with all edges either pointing toward the root or away from it.

They use the \textit{maximum influence paths (MIP)} to estimate the influence from one node to another. Let $\pi(G,v,v')$ be the set of all paths from $v$ to $v'$ in $G$, $P=\langle v, v_i, v_{i+1}, \ldots, v' \rangle$ be a path and $p(P) = \prod_{i=1}^{m-1}p(v_i,v_{i+1})$ be its propagation probability. Then the MIP can be formally defined as follows.

\begin{definition}[Maximum influence path]
	The maximum influence path for a graph $G$ from node $v$ to node $v'$ is defined as:
	\begin{equation*}
		MIP_G(v,v') = \argmax_{P} \{p(P) | P \in \pi(G,v,v')\}.
	\end{equation*}
\end{definition}

Notice that $MIP_G(v,v')$ is always unique, and any subpath in $MIP_G(v,v')$ from $i$ to $j$ is also the $MIP_G(i,j)$.

Thus, first MIPs between every pair of nodes are computed by adopting the Dijkstra algorithm~\cite{dijkstra1959note}, ignoring MIPs with probability smaller than some influence threshold $\theta$.
Then, they union the MIPs starting or ending at each node into the arborescence structures, which represent the local influence regions of each node. 
The authors consider influence propagated through these local arborescences, and we refer to this model as the \textit{Maximum Influence Arborescence (MIA)} model.
The influence spread in the MIA model is submodular, and so the usual greedy algorithm guarantees an influence spread within $\left(1 - \frac{1}{e} \right)$ of the optimal solution.
Experimentally compared with the other described heuristics, MIA algorithm is always among the best, and in most cases it significantly outperforms the rest heuristics, with a margin as much as 100-260\% influence spread. Moreover, the authors show that by tuning $\theta$, it is possible to adjust the trade-off between efficiency (in terms of running time) and effectiveness (in term of influence spread).

\paragraph{CELF++}
Another algorithm has been proposed in~\cite{goyal2011celf}, namely CELF++, which improves CELF by inserting additional heuristic optimizations exploiting submodularity.
It is empirically 17-34\% faster than CELF.

\paragraph{IRIE}
In~\cite{jung2012irie} a novel algorithm that outperforms MIA has been designed, \textit{IRIE}. 

The rationale behind this new algorithm is a novel Influence Ranking method, IR, derived from a belief propagation approach, which uses a small number of iterations to generate a global influence ranking of the nodes and then select the highest ranked node as the seed. 
However, the influence ranking is only good for selecting one seed. If we use the ranking to directly select $k$ top ranked nodes as $k$ seeds, their influence spread may overlap with one another, and not result in the best overall influence spread.
To overcome this issue, the authors integrate IR with a simple Influence Estimation, IE, such that,
after one seed is selected, they estimate additional influence impact of this seed to each node in the network, which is much faster than estimating marginal influence for many seed candidates, and then use the results to adjust next round computation of influence ranking. 
When combining IR and IE together, IRIE is obtained.

An interesting result for IR when applied on tree graphs is the following.
For each edge $(v,v') \in E$, let us denote with $m(v,v')$ the expected number of activated nodes when $S = \{v\}$ and $(v,v')$ is removed from $E$. Let $\tilde{\sigma}(v)$ and $\tilde{m}(v,v')$ be the estimates for $\sigma(v)$ and $m(v,v')$, respectively.
Then, we compute the estimates as follows:

\begin{equation*}
	\tilde{\sigma}(v) = 1 + \sum_{v' \in N^{out}(v)} p(v,v') \cdot \tilde{m}(v',v)
\end{equation*}

\begin{equation*}
	\tilde{m}(v,v') = 1 + \left(\sum_{w \in N^{out}(v),w \ne v'} p(v,w) \cdot \tilde{m}(w,v)\right)
\end{equation*}

The following theorem holds.

\begin{theorem}
	For any tree graph, for each node $v$, $\tilde{\sigma}(v) = \sigma(v)$ and, for each edge $(v',v \in E), \tilde{m}(v,v') = m(v',v)$. 
\end{theorem}

Finally, IRIE has another additional feature, namely, it works also on the IC-N diffusion model, i.e., the Independent Cascade model in which also negative opinions can spread among the nodes.

\paragraph{Quasi Linear Time Algorithm}
In~\cite{borgs2014maximizing}, the authors provide a significant speed-up in computing a solution for IMP under the IC diffusion model, with an approximation factor of $\left(1 - \frac{1}{e} - \epsilon\right)$, achieving a final complexity of $O(kl^2(m+n)\log^2 n \epsilon^{-3})$ w.r.t. $\Omega(mnk\cdot Poly(\epsilon^{-1}))$ achieved in~\cite{chen2009efficient}. Recall that $n=|V|, m=|E|$, $k$ is the budget for the seeds and $l$ is some integer factor.
The proposed algorithm is randomized and its basic version succeeds with probability equal to $0.6$, and, since failure is detectable, such a probability can be increased by repetition.

To understand the approach underlying the algorithm, let us start by considering the problem of finding the single node with highest influence. 
Consider the following \textit{polling} process: extract randomly some node $v'$ with uniform probability, and determine the set of nodes that would have influenced $v'$. 
Intuitively, if we repeat this process multiple times, and a certain node $v$ appears often as an \textit{influencer}, then $v$ is likely a good candidate for the most influential node. 
In fact, the authors show that the probability a node $v$ appears in a set of influencers is proportional to the expected number of nodes that will be influenced by $v$, and standard concentration bounds show that this probability can be estimated accurately with relatively few repetitions of the polling process. Moreover, it is possible to efficiently find the set of nodes that would have influenced a node $v'$: this can be done by simulating the influence process, starting from $v'$, in the transpose graph (i.e., the original network with edge directions reversed).
Thus, the algorithm proceeds in two steps. 
First, it repeatedly applies the random sampling technique described above to generate a sparse hypergraph representation of the network. Each hypergraph edge corresponds to a set of individuals that was influenced by a randomly selected node in the transpose graph. 
This hypergraph encodes our influence estimates: for a set of nodes $S$, the total degree of $S$ in the hypergraph is approximately proportional to the influence of $S$ in the original graph. 
In the second step, the standard greedy algorithm is run on the hypergraph to return a set of size $k$ of approximately maximal total degree.
The final complexity achieved by the algorithm is due to the running time required by the first phase.~\footnote{The formula for the complexity of the algorithm reported above is the one corresponds to the version of the algorithm for which the probability success has been amplified beyond $0.6$.}

\paragraph{TIM}
Starting from these results, in~\cite{tang2014influence} an even faster algorithm to solve IMP under IC~\footnote{Besides working when IC is adopted as diffusion model, TIM also supports the Triggering Diffusion model.} is designed, achieving a complexity equal to $O((k+l)(n+m)\log \epsilon^{-2})$, returning a solution with an approximation factor of $\left(1 - \frac{1}{e} - \epsilon\right)$ with at least $1-n^l$ probability.
Borrowing some ideas from the algorithm of~\cite{borgs2014maximizing}, TIM consists in two main parts. 
First, there is the \textit{parameter estimation} phase, which computes a lower-bound of the maximum expected spread among all size-$k$ node sets, and then uses the lower-bound to derive a parameter $\theta$.
Then, there is the \textit{node selection}, which samples $\theta$ random $RR$ sets from the graph $G$, where an $RR$ set is a reverse reachable set, defined as follows.

\begin{definition}[Reverse reachable set (RR)]
	Let $v$ be a node in $G$, and $g$ be a graph obtained by removing each edge $e$	in $G$ with $1 - p(e)$ probability. The reverse reachable $(RR)$ set for $v$ in $g$ is the set of nodes in $g$ that can reach $v$.
\end{definition}

This way, a size-$k$ node set $S^*$ that covers a large number of $RR$ sets is derived and then returned as as the final result.

\paragraph{SKIM}
Still willing to achieve a higher scalability, in~\cite{cohen2014sketch} another approximation algorithm for IMP under IC is proposed.
SKIM (SKetch-based Influence Maximization) algorithm works on per-node summary structures, called \textit{combined reachability sketches}.
The sketch of a node compactly represents its influence \textit{coverage} across instances. The combined
reachability sketch of a node is the \textit{bottom-$k$ min-hash sketch}~\cite{cohen2015all} of the combined reachability set of the node, thus generalizing the reachability sketches of Cohen~\cite{cohen1997size}, which have been defined for a single instance. 
The rationale behind such a concept is that we want to capture the reachability information of the nodes across different instances.
Here, the parameter $k$ is a small constant that determines the trade-off between computation and accuracy. Bottom-$k$ sketches of sets support cardinality estimation, which means that we can estimate the influence of a node or of a set of nodes from their combined reachability sketches. 
SKIM scales by running the greedy algorithm in \textit{sketch-space}, always taking a node with the maximum estimated (rather than exact) marginal contribution.
SKIM computes combined reachability sketches, but only until the node with the maximum estimated influence is computed. This node is then added to the seed set. 
Next, the sketches are updated w.r.t. a residual problem in which the node that is selected into the seed set and its \textit{influence} are no longer present. 
SKIM resumes the sketch computation, starting with the residual sketches, but stopping when a node with maximum estimated influence is found. 
A new residual problem is then computed. This process is iterated until the seed set reaches the desired size. Since the residual problem becomes smaller with iterations, we can compute a very large seed set very efficiently.

Another important contribution is the introduction influence oracles: after preprocessing that is
almost linear, we can answer influence queries very efficiently, considering only the sketches of the query seed set.
Specifically, for instances with $n$ nodes and $m$ edges, the sketches are built in $O(km)$ total time. The influence of a set $S \subseteq V$ can then be approximated from the sketches of the nodes in $S$.
The oracle applies the union cardinality estimator\cite{cohen2009leveraging} to estimate the union of the influence sets of the seed nodes. 
The query runs in time $O(k|S|\log |S|)$ and unbiasedly with a well-concentrated relative error of $\epsilon = \frac{1}{\sqrt{k}}$.
Notice that, while preprocessing depends on the number of instances, the sketch size and the approximation quality only depend on the sketch parameter $k$.

\subsubsection{Linear Threshold model}

\paragraph{SPIN}
In~\cite{narayanam2011shapley}, the authors propose a Shapley value based heuristic, ShaPley
value based Influential Nodes (SPIN), for the LT model. However, SPIN only relies on the evaluation of influence spreads of seed sets, and thus does not use specific features of the LT model. Moreover, SPIN is not scalable.

\paragraph{LDAG}
A significant step has been taken in~\cite{chen2010lt}, where it is shown that computing influence in Directed Acyclic Graphs (DAGs) can be done in time linear w.r.t. the size of the graphs. 
Because of the fast computation in DAGs, the authors propose the first scalable influence maximization algorithm designed for the LT model.

First, they show that computing the exact influence spread in the LT model is $\#P$-hard, even if there is only one seed in the network. 
This hardness result is important since it closes the open problem left in~\cite{kempe2003maximizing} and
further indicates that the greedy algorithm may have intrinsic difficulty to be made more efficient. 
Then, it is shown that computing influence spread in DAGs can be done in linear time, which relies on an important linear relationship in activation probabilities between a node and its in-neighbors in DAGs.
Next, based on the fast influence computation for DAGs, the authors propose the first scalable heuristic algorithm tailored for influence maximization in the LT model, namely LDAG.
The rationale is to construct a local DAG surrounding every node $v$ in the network, and restrict the influence to $v$ to be within the local DAG structure. This makes influence computation tractable and fast on a small DAG. 
To select local DAGs that could cover a significant portion of influence propagation, we propose a fast greedy algorithm adding nodes into the local DAG of a node $v$ one by one, such that the individual influence of these nodes to $v$ is larger than a threshold parameter. 
After constructing the local DAGs, we combine the greedy approach of selecting seeds that provide the maximum incremental influence spread with a fast scheme of updating incremental influence spread of every node. The combined fast local DAG construction and the fast incremental influence update make the LDAG algorithm very efficient.
Experimental results show that our LDAG algorithm scales to networks with millions of nodes and edges, while the optimized greedy algorithm already take days for networks in the size of 64K. In term of influence spread, LDAG algorithm is always very close to that of the greedy algorithm in all test cases, showing that it is able to achieve the same level of influence spread while running
in orders of magnitude faster than the greedy algorithm.

\subsubsection{Dynamic Diffusion model}

\paragraph{FastMargin}
In~\cite{xie2015dynadiffuse}, the authors prove that IMP under DynaDiffuse is NP-hard. 
Moreover, they prove that, also in this model, the influence function $\sigma(\cdot)$ is monotonic and submodular, thus the standard greedy algorithm provided achieves the usual approximation factor of $\left(1 - \frac{1}{e}\right)$. However, here a faster algorithm is provided. 
First it predicts a relatively precise influence spread using stochastic model checking on the CTMC Parallel Composition.
Then, it predicts the increased marginal influence spread $\sigma(I \cup \{v\} )- \sigma(I)$ of adding each $v \in V \setminus I$ when selecting an additional node to be added to the current initial set.
To improve the efficiency, they estimate this marginal influence spread using a faster discounted formula for $\sigma(\{v\})$ instead of determining a precise value with stochastic model checking.

For scalability to large networks, they also adopt the CELF lazy evaluation approach~\cite{leskovec2007cost}, which dramatically reduces the number of evaluations of $\argmax_v(\sigma(I \cup \{v\}) - \sigma(I))$.
Such an algorithm can be flexibly used in any continuous-time constrained diffusion model that assumes that the diffusion rate follows the exponential distribution.
The total complexity is $O(|V||\zeta|^2+(k-1)|V|d^*)$, where $d^*$ is the maximum degree of all nodes and $\zeta$ is the set of states.

\subsubsection{Heat Diffusion model}

\paragraph{CIM}
Adopting the Heat Diffusion Model, in~\cite{chen2014cim} a novel technique is employed to detect the most promising nodes that will maximize the spread throughout the network. They propose a three-level approach based on the idea of \textit{community}. The three phases of Community-based Influence Maximization (CIM) are community detection, candidate generation, and seed selection. Let us analyze them.

First, the authors argue that nodes in social networks naturally cluster together. To reduce the overhead incurred in computing influence spreads based on HDM, they explore the clustering phenomenon among nodes in a community to avoid the computation of overlapped influence spreads among nodes in the same community.
In other words, once a node in a community is selected, it is very unlikely that selecting another node belonging to the same community results being better than placing a seed in a node from a different one.

For the clustering algorithm, H\_Clustering, they incorporate the notion of modularity,  as expressed in~\cite{newman2006modularity}, based on a bottom-up approach to iteratively merge nodes with strong structure similarity into communities. 
First, for each node in the given social network, the algorithm derives the structural similarity between the node and its neighboring nodes, where the structure similarity is used as the edge weight for its
neighboring nodes.
Then, each node is considered as a community, and groups each pair of nodes into a community if the structural similarity between these two nodes is the largest among their surrounding edges from each other. 
In other words, given two nodes $v$ and $v'$, if the edge $(v,v')$ is the largest among all edges connecting to $v$ and also is the largest among all edges connecting to $v'$, merge $v$ and $v'$ into a community. 
Next, each newly created community becomes a node, and the process continues until a termination condition
is reached. 
Such a termination condition should measure the quality of discovered communities in order to decide when to stop the community detection process.
This is why, as done in~\cite{huang2010shrink}, they adopt \textit{modularity gain}~\cite{feng2007novel}. 
To reduce the search space of seed nodes, one way could be analyzing the community structures and prune off insignificant communities and their nodes.
Finally, even though we may want to avoid placing multiple seeds in the same community, the size of community is still a factor for seed placement, i.e., communities should not be treated the same since  placing seeds in a large community could trigger more nodes than in a small community.

To select the seeds from the large communities just built, the centroids of communities may appear as the natural candidates. However, also hubs, i.e., nodes connecting different communities, should be carefully considered.
A community is called \textit{significant} if it has have the number of nodes larger than the average number of nodes a seed may influence in a given influence maximization task.
In order to generate a small set of seed candidates, he approach is to consider only the potential centroid nodes (i.e., nodes with high degree and large score sum) in significant communities and the hub nodes as candidates by eliminating outliers and nodes in some \textit{insignificant communities}. Then, the top-$p\%$ of high-degree nodes and large-score-sum nodes in each significant community and all hub nodes connecting significant communities as the candidate set.

Finally, we should select $k$ seeds out of all the candidates.
For various combinations of $k$ seeds selected from the candidate seed set, the computation of influence spreads is carried out by simulating how influence spreads from those seeds based on the heat diffusion model.
To overcome the computational issues that arise while performing this task, the authors designed a two-step approach.
First, a quota-based method is employed to determine the number of seeds to be allocated for a given significant community and then determine which nodes should be selected for this significant community based on a heuristic of position score and hub purity. 
Then, in the second step, CIM heuristically finds a new candidate seed which may potentially increase the influence spread to swap with a seed node, aiming to obtain a better seed set, basically applying a local search approach. 
This process is repeated until the influence spread does not improve any more or a certain number of seed node swapping has been performed.

\subsection{Seeds Placement}\label{sec:results_seeds}
We now turn to the results concerning new features that have been added to IMP w.r.t. the seeds placement.

\paragraph{FindHighDegreeVertex}
We restrict possible operations that may be performed on the graph only to \textit{jump} and \textit{crawl}, adopting such a number as the complexity measure. Recall that the goal is to find interesting individuals in the network both in terms of degree and CC.

Given $0 < \beta < 1$, we want to find a vertex $v$ such that: $d^* \leq d(v)n^{1-\beta}$, with $d^*$ being the maximum degree of the different nodes and $n=|V|$. 
To find such a vertex, one path could be the following: if $d^*$ is smaller than $n^{1-\beta}$, then any vertex would satisfy our condition.
If this does not hold, we observe that the expected size of a random sample we need to take to obtain a neighbor with maximum degree is $\frac{n}{d^*}$. Thus, we can just sample $\frac{n}{d^*}$ nodes: if one has a degree higher than $\frac{d^*}{n^{1-\beta}}$, otherwise such a vertex is the neighbor of a node with degree $d^*$.

\textit{FindHighDegreeVertex} works assuming to know the value of $d^*$: to remove such a constraint, we just simulate the possible values of $d^*$ in a logarithmic fashion.
For such an algorithm, the following result holds.

\begin{theorem}
	For any $0 < \beta < 1$, \textit{FindHighDegreeVertex} approximates the maximum degree to an expected multiplicative factor of $O(n^{1-\beta})$ performing $n^{\beta}\log n$ \emph{jump} and \emph{crawl} operations.
\end{theorem}

Even more interestingly, the authors show that the algorithm is optimal up to a logarithmic factor, as stated in the next result.

\begin{theorem}
	For any $0 < \beta < 1$, if some algorithm $\mathcal{A}$ performs at most $n^{\beta}$ jump and crawl operations, then $\mathcal{A}$ approximates the maximum degree to an expected multiplicative factor of $\Omega(n^{1-\beta})$.
\end{theorem}

Next, we consider graphs with the power law property since the degree distribution of many networks resembles such a behavior.
Intuitively, by power law we mean that the fraction of vertices with degree $d$ is \textit{close} to $\frac{1}{d^{\gamma}}$, with $\gamma > 1$ and $d$ \textit{big enough}. We can state the following.

\begin{theorem}
	Let $0 < \beta < \frac{\gamma -1 }{\gamma}$, with $\gamma > 2$. Then \textit{FindHighDegreeVertex} can be adapted so that it performs  $O(n^{1-\beta})$ jump and crawl operations, approximating the maximum degree to an expected multiplicative factor of $O(n^{\frac{1}{\gamma}-\frac{\beta}{\gamma-1}})$.
\end{theorem}

The last step is considering networks created by the preferential attachment process of Barabasi~\cite{barabasi1999emergence}. 
Adapting \textit{FindHighDegreeVertex} for such cases, exploiting the idea of lazy random walk,\footnote{A lazy random walk on a connected graph $G$ prescribes to stay in the current vertex with probaility $0.5$ and to move to a uniformly chosen random neighbor with probability $0.5$. Such a random walk forms an ergodic Markov chain.} the following result holds.

\begin{theorem}
	Let $0 < \beta < \frac{1}{11}$. Then, there exists a modification of \texttt{FindHighDegreeVertex} that approximates $d^*$ up to an expected multiplicative ratio of $O(n^{\frac{1}{2}-\beta})$ performing $O(n^{\beta}\log n)$ jump and crawl operations.
\end{theorem}

We are now ready to state the last two results.

\begin{theorem}
	For any $0 < \beta < 1$, there exists an algorithm returning a $\left(\frac{1}{\log n},n^{\beta},\frac{1}{\log n} \right)$ approximation to maximum CC, performing $\tilde{O}(n^{1-\beta})$ operations.~\footnote{We recall that $\tilde{O}(\cdot)$ corresponds to the usual \textit{big-O} notation but ignoring any logarithmic factor and that the mixing time of a Markov chain is the time until the Markov chain is \textit{close} enough to its steady state distribution.}
\end{theorem}

\begin{theorem}
	For any $0 < \beta < 1$, in networks following a power law with $\gamma \geq 3$ and such that the lazy random walk mixes in time $\tau$, there exists an algorithm returning a $\left(1,n^{\beta},\frac{1}{\log n} \right)$ approximation to maximum CC, performing $\tilde{O}(n^{1-2\beta})$ operations.
\end{theorem}

\paragraph{AdaptiveSeeding}
The main idea is to have access to a small number of nodes, and then explore only the neighbors of such nodes, which can be accessed with some probability after one of their neighbor is activated. Moreover, at most $k$ seeds can be placed in total~\cite{seeman2013adaptive}. 
Let us start by tackling problem of stochastic optimization, i.e., which should be the first nodes to be selected as starting seeds.

One of the most common techniques in the literature is the Sample Average Approximation method~\cite{kleywegt2002sample}, whose rationale is to sample instances from the product distribution over the set of neighbors and then optimize the influence function w.r.t. such sampled case.
If such a function is simple, e.g., additive, then such an approach could be directly extended and when the function is also submodular, the $\left(1-\frac{1}{e}\right)$-approximation result proved in~\cite{nemhauser1978analysis} holds.

However, if $\sigma(\cdot)$ is complex, the only solution is designing a non adaptive approach, which commits to the $k$ nodes before both stages are realized. Of course, such policies will be significantly weaker than the adaptive ones. 
The algorithmic challenge arises in attempting to optimize the influence function of the Triggering model. Interestingly, any influence function in the Triggering model can be adaptively seeded, i.e., there is an algorithm which finds an adaptive policy that is a constant factor approximation to the optimal adaptive policy, for any influence function in this class. 
The main idea is to design an algorithm which approximates the optimal randomized-and-relaxed non adaptive policy and show that its solution is a good approximation to the optimal adaptive policy.
First a concave function is constructed and the problem is formulated as optimizing the concave function
under a mixed constraint of integral and fractional linear constraints and show that solutions to this objective are an upper bound on adaptive policies.

Then, the authors present an algorithm which mimics a gradient-ascent process, taking steps in the direction of the \textit{densest gradient}, showing such an algorithm obtains a constant factor approximation to the optimal adaptive policy.

\subsection{Network Uncertainties}\label{sec:results_network}
Finally, we look at the main results obtained when considering IMP formulated on networks for which we do not have all the information.

\paragraph{IMUG}
In~\cite{mihara2015influence}, we are given the number of nodes constituting the network, but we do not have any information either on the existence or on the probabilities associated to the different edges. The goal here is to reconstruct a network that is as close as possible to the real connections among the individuals, and then solve IMP. 
There are $R$ rounds, and in each round we can probe $m$ nodes, select $k$ seed nodes, and trigger influence spread from the selected $k$ seeds. 
The selected seed nodes become activated, spread influence to their adjacent node, and each newly activated (influenced) node recursively repeats the influence spread process
according to a given diffusion model.~\footnote{The authors provide an algorithmic scheme that can be adopted with both the IC and LT diffusion models.}
Probing node $v$ obtains a list of nodes adjacent to $v$. In each round, we can repeat this probing process $m$ times. We can obtain a list of activated nodes in each round. 

The proposed heuristic algorithm is called Influence Maximization for Unknown Graphs (IMUG). 
The rationale behind it is to greedily probe the node with the highest expected degree from unprobed nodes, and greedily select the inactive node with the highest expected degree as a seed node by using the results of past probing and influence spread. 
In these cases, we have to rely on available local information about node degrees, being not possible to obtain complete knowledge of the entire topological structure of the graph.
Using node degree is a straightforward solutions in situations of limited information since high-degree nodes tend to spread more influence than low-degree nodes do, and the effectiveness of degree-based heuristic algorithms have been successfully shown~\cite{chen2009efficient}.
Note that the degree of each node is not available in the problem studied here, and therefore it is estimated from the results of past probing.

As a probing strategy, the authors adopt a biased sampling strategy called Sample Edge Count (SEC), also known as a snowball sampling strategy~\cite{maiya2011benefits}. 
SEC greedily probes the node with the highest expected degree. Given a set of already probed nodes $D$, SEC estimates the expected degree of node $v$ in the original graph $G$ as the degree of node $v$ in the induced subgraph of $D \cup \{v\}$, and probes the node with the highest expected degree. 
For each round, we repeat SEC probing $m$ times.

In the initial state, IMUG estimates each node expected degree $d(v)$, which are IMUG parameters. 
If some knowledge of node degrees, such as the average degree, is available in advance, we can determine $d(v)$s by using that knowledge. When we have no information about node degree, a simple option is to set $d(v)$ to $0$.
In each round, IMUG updates the expected degree of each node by using the results of SEC probing. When node $v$ is probed, since the true degree of node $v$ is known, it is fixed to the known true degree.
Moreover, for each node $v'$ adjacent to node $v$, since node $v'$ is revealed to have at least one link, $d(v')$ is incremented by one unless the degree of node $v'$ is fixed.

Despite being a heuristic, IMUG achieves a 60-90\% influence spread as compared with algorithms using the entire social network topology even when only 1-10\% of such a topology is known.

\paragraph{ARISEN}
In~\cite{wilder2018maximizing}, ideas from both~\cite{brautbar2010local} and~\cite{mihara2015influence} are combined. The authors propose the \textit{exploratory influence maximization}, looking for an algorithm that makes a small number of queries and returns a set of seed nodes which are approximately as influential as the globally optimal seed set.
The approach is not comparable with~\cite{mihara2015influence} since the algorithm observes all of the nodes which are activated by its chosen seeds, which can reveal a great deal about the network.
As in~\cite{brautbar2010local}, they adopt the \textit{jump} and \textit{crawl} operations, when visited, a node reveals all of its edges, the query cost of an algorithm is the total number of nodes visited using either operation and the goal is to find influential nodes with a query cost that is much less than the total number of nodes.

The graphs adopted for he analysis are drawn from the \textit{Stochastic Block Model (SBM)}, which originated in sociology~\cite{fienberg1981categorical}. 
In the SBM, the network is partitioned into disjoint communities $C_1, \ldots, C_l$.
Each within-community edge is present independently with probability $p_w$ and each between-community edge is present independently with probability $p_b$. 
Each community $C_i$ is internally drawn as $\mathcal{G}(|C_i|,p_w)$ with additional random edges to other communities. 

ARISEN (Approximating with Random walks to Influence a Socially Explored Network) takes as input the parameters $n, p_w, p_b$, but is not given any prior information about the realized draw of the network. 
The rationale behind ARISEN is to sample a set of $T$ random nodes $\{v_1,\ldots,v_T\}$ from $G$ and explore a small subgraph $H_i$ around each $v_i$ by taking R steps of a random walk.  $R$ and $T$ are inputs: $T$ should be greater than $k$ so we can be sure of sampling each of the largest $K$ communities.
The subgraphs $H_i$ are used to construct a weight vector $w$ where $w_i$ gives the weight associated with $v_i$.  The algorithm then independently samples each seed from $\{v_1,\ldots,v_T\}$ with probability proportional to $w$.

ARISEN uses the random walk around each $v_i$ to estimate the size of the community that $v_i$ lies in. From these estimates, it constructs a $w$ that, in expectation, seeds the largest $k$ communities. Then, it tests if a $w'$ that puts more weight on large communities would increase the expected influence.

\paragraph{CMAB for IM}
In~\cite{vaswani2015influence}, the authors assume that the influence probabilities are not known and it is not possible to estimate them just by data. This is why they propose to learn them \textit{online}, employing CMAB. 
The authors study the problem adopting the IC, even though the provided framework is valid for any discrete time diffusion model. The actual spread is the number of nodes reachable from the selected seed nodes in the true possible world and we denote it by $\bar{\sigma}$.
The mapping between CMAB and IMP is the following:
\begin{itemize}
	\item associate an arm $i$ to each edge $e \in E$;
	\item reward $X_{i,t}$ represents the status of the edge, which can be either live, i.e., an attivation attempt along it succeeded, or dead;
	\item mean $\mu_i$ is the weight of the edge corresponding to arm $i$;
	\item superarm A is associated to the union of outgoing edges $E_S$ from the seeds $s \in S$;
	\item reward $r_t$ in round $t$ corresponds to the spread $\sigma$ in the $t$-th diffusion attempt.
\end{itemize}

In each round, the regret minimization algorithm selects a seed set $S$ cardinality $k$ and plays the corresponding superarm $E_S$. 
$S$ can be selected either randomly or by solving IMP with the current influence probability estimates 
Oracle $O$ takes as input the graph $G$ the estimates of the means, and outputs a seed set $S$. 
For the case of IMP, the oracle constitutes a $\left(1-\frac{1}{e},\frac{1}{|E|}\right)$-approximation oracle~\cite{chen2016combinatorial}. 

Once the superarm is played, information diffuses in the network and a subset of network edges become live which leads  to a subset of nodes becoming active. 
The reward $X_{i,s}$ for these edges is $1$. 
The reward $\sigma(S)$ is the number of active nodes at the end of the diffusion process and is thus a non-linear function of the rewards of the triggered arms. 
After observing a diffusion, the mean estimate $\hat{\mu}$ needs to updated. 
In this context, the notion of a feedback mechanism plays an important role since it characterizes the information available after a superarm is played. 
This information is used to update the model to improve the mean estimates. 
Let $S^*$ be the solution and let $\sigma^* = \sigma(S^*)$ the optimal expected spread. 
Since IMP is $\mathsf{NP}$-hard, even if the true influence probabilities are known, we can only hope to achieve an expected spread of $\alpha \beta \sigma^*$, where $\alpha = 1-\frac{1}{e}$ and  $\beta = \frac{1}{|E|}$.
Call $S_s$ the seed set chosen by $A$ in round $s$. The regret incurred by $A$ is then defined by:

\begin{equation*}
	R(\mu,\alpha,\beta) = T\alpha \beta \sigma^* - \mathbb{E}_S\left[\sum_{s=1}^T \sigma(S_s)\right]
\end{equation*}

where the expectation is over the randomness in the seed sets output by the oracle.
The usual feedback mechanism is the edge-level feedback proposed by~\cite{chen2016combinatorial}, where we assume that we know the status of each triggered edge in the \textit{true} possible world. 
The mean estimates of the arms distributions can then be updated as follows:

\begin{equation*}
	\hat{\mu}_i = \frac{\sum_{s=1}^t X_{i,s}}{T_{i,t}}.
\end{equation*}

However, edge-level feedback is often not realistic because success/failure of activation attempts is not generally observable. Unlike the status of edges, it is quite realistic and intuitive that we can observe the status of each node.
While this is a more realistic assumption, the disadvantage of the node-level feedback is that updating the mean estimate for each edge is more challenging. 
This is because we do not know which active parent activated the node, or when it was activated. 
Under edge-level feedback, we assume that we know the status of each edge starting from the neighbors of some node $v$, and use it to update mean estimates. 
Under node-level feedback, any of the active parents may be responsible for activating a node $v$ and we don’t know which. 
The authors provide two ways to solve this issue.

The most common way to infer the edge probabilities given the status of each node in the cascade is to
use Maximum Likelihood Estimation (MLE). They use an MLE formulation similar to those proposed in~\cite{netrapalli2012learning,saito2008prediction}. 
These works describe an offline method for learning influence probabilities, where a fixed set of past diffusion cascades is given as input. The log-likelihood function for a given set of cascades $C$ is given by:

\begin{equation*}
	\log L(\overrightarrow{p}) = \sum_{c=1}^C \sum_{v \in V} log L_{v}^c(\overrightarrow{p})
\end{equation*}

where $L_{v}^c(\overrightarrow{p})$ models the likelihood of observing the cascade $c \in C$ w.r.t. node $v$, given the influence probability estimates $\overrightarrow{p}$. 
The authors then provide a bound,~\footnote{For more details, please see in~\cite[Theorem 1]{vaswani2015influence}} which limits the price we pay in terms of error for adopting the node-level feedback over edge-level feedback mechanism.

Unfortunately, the time complexity of the this approach is $O(|E|T^2)$, which does not scale to networks with a large number of edges. 
To mitigate this, the authors adapt a result from online convex optimization for learning the edge probabilities, in which an online convex optimization framework has been developed to minimize a sequence of convex functions over a convex set~\cite{zinkevich2003online}. 
In such a case, they solve an online convex optimization problem for each node in the network. 
It is shown that with sufficiently many rounds $T$, the parameters learned by the online MLE algorithm are nearly as good as those learned by the offline algorithm. 

The other approach is the following.
In typical social networks, the influence probabilities are very small and this causes the number of active parents to be small, too. 
The authors propose a scheme where we choose one of the active neighbors of $v'$, say $v$, uniformly at random, and assign the credit of activating $v'$ to $v$. 
The probability of assigning credit to any one of $k$ active parents is $\frac{1}{k}$.
In other words, edge $(v,v')$ is given a reward of 1 whereas edges corresponding to other active parents  are assigned a zero reward. 
Then the same approach of the edge-level feedback is adopted.
Observe that we could make a mistake by inferring an edge to be live while it is dead in the true world or vice versa. 
We term the probability of such faulty inference the \textit{failure probability under node-level
	feedback}. An important question is whether we can bound this probability. This is important since failures could ultimately affect the achievable regret and the error in the learned probabilities. 
As the number of active nodes for a cascade increases, the error in the mean estimates increases and it is better to use the maximum likelihood approach for credit distribution. The authors empirically find that the proposed node-level feedback achieves competitive performance compared to edge-level feedback.

The possible learning algorithms that can be adopted to instantiate the described approach are Combinatorial Upper Confidence Bound(CUCB)~\cite{chen2013combinatorial}, $\epsilon$-Greedy~\cite{chen2013combinatorial}, Thompson Sampling~\cite{bubeck2012regret} and Pure Exploitation.

\paragraph{LUGreedy}
The second way of dealing with uncertainties on social networks is by studying a robustness framework. 
In~\cite{chen2016robust}, for RIMP, the knowledge of the confidence interval assumed to be the input. 
When the amount of observed information cascade is small, the best robust ratio $\max_S g(\Theta,S)$ for the given $\Theta$ can be low so that the output for a RIM algorithm does not have a good enough guarantee of the performance in the worst case. 
Then a natural question is the following: given $\Theta$, how to further make samples on edges so that $\max_S g(\Theta,S)$ can be efficiently improved?

Consider RIM, $\Theta=\times_{e \in E}[l_e,r_e]$, the true probability $\theta \in \Theta$ to be unknown. Let us denote $\theta^- = (l_e)_{e \in E}$ and $\theta^+ = (r_e)_{e \in E}$. We have the following result.

\begin{theorem}
	RIMP is $\mathsf{NP}$-hard and, for any $\epsilon > 0$, it is $\mathsf{NP}$-hard to find a seed set $S$ with a robust ratio at least equal to $1 - \frac{1}{e} + \epsilon$.
\end{theorem}

Then, the authors propose Lower-Upper Greedy (LUGreedy) algorithm, which outputs the best seed set $S_{\Theta}^{LU}$ for $\theta^-$ such that:

\begin{equation*}
	S_{\Theta}^{LU} = \argmax_{S \in \{S_{\theta^-}^g,S_{\theta^+}^g\}} \{\sigma_{\theta^-}(S)\}
\end{equation*}

To evaluate the performance of this output, we first define the gap ratio $\alpha(\Theta) \in [0,1]$ of the input parameter space to be:

\begin{equation*}
	\alpha(\Theta) = \frac{\sigma_{\theta^-}(S_{\Theta}^{LU})}{\sigma_{\theta^+}S_{\theta^+}^{g}}
\end{equation*}

Then, LUGreedy achieves the following result.

\begin{theorem}
	Given a	graph $G$, parameter space $\Theta$ and budget limit $k$, LUGreedy	outputs a seed set $S_{\Theta}^{LU}$ of size $k$ such that:
	
	\begin{equation*}
		g(\Theta,S_{\Theta}^{LU}) \geq \alpha(\Theta) \left(1 - \frac{1}{e}\right).
	\end{equation*}
\end{theorem}

Notice that the worse-case bound could be small if $\Theta$ is not assumed to be tight enough.
Moreover, the best possible robust ratio $g(\Theta,S)$ can be too low so that the output for RIMP could not provide us with a satisfying seed set in the worst case. 
However, according to the Chernoff's bound, the more samples we make on an edge, the narrower the confidence interval we get that guarantees the true probability to be located within the confidence interval with a desired probability of confidence. 
Thus, after sampling to get a narrower parameter space, we could use LUGreedy algorithm to get the seed set.
The authors propose two ideas exploiting properties of additive and multiplicative confidence interval respectively to this issue and incorporate into Uniform Sampling algorithm.

The first idea is that we may sample every edge for sufficient times to shrink their confidence intervals in $\Theta$, and feed LUGreedy with $\Theta$ as same as solving RIMP, then the performance is guaranteed by the factor we found for LUGreedy. 
The second idea is to use the multiplicative confidence interval to reduce the fluctuation of influence spread, then LUGreedy still applies. The ratio of influence spread can be bounded based on the relation of $l_e$ and $r_e$ in the multiplicative form.
To unify both ideas mentioned above, they propose Uniform Sampling for RIM algorithm, which samples every edge with the same number of times, and use LUGreedy to obtain the seed set.

Finally, the authors deal with instances in which the influence probabilities are different. 
The intuition for non-uniform sampling is that the edges along the information cascade of important seeds determine the influence spread, and thus they should be estimated more accurately than other edges not along important information cascade paths. 
Starting from the seed set $S$, once node $v$ is activated, it will try to activate its out-neighbors. That is, for every out-edge $e$ of $v$, denote $\eta_e$ as the number of samples, then $e$ will be sampled once to generate a new observation $x_e^{\eta_e}$ based on the latent Bernoulli distribution with success probability $p_e$, and $\eta_e$ will be increased by $1$. 
The process goes on until the end of the information cascade. 
To solve such a problem, the authors propose Information Cascade Sampling for RIM algorithm, which adopts information cascade sampling described above to select edges.

\paragraph{SaturateGreedy}
The other sense in which the word robust has been used is w.r.t. the guarantees that an algorithm may provide w.r.t. a spectrum of different diffusion models and parameter settings~\cite{he2016robust}. 
To model this uncertainty, the authors assume that the algorithm is presented with a set $\Sigma$ of influence functions, and assured that one of these functions actually describes the influence process, but not told which one. The set $\Sigma$ could be finite or infinite. 
Since the algorithm does not know $\hat{\sigma}$, it must simultaneously optimize for all objective functions in $\Sigma$, in the sense of maximizing:

\begin{equation*}
	\rho(S_0) = \min_{\hat{\sigma} \in \Sigma} \frac{\hat{\sigma}(S_0)}{\hat{\sigma}(S^*)}
\end{equation*}

where $S^* \in \argmax_S \hat{\sigma}(\cdot)$ is an optimal solution knowing which function $\hat{\sigma}(\cdot)$ is to be optimized. 

First, the authors show that, unless the algorithm gets to exceed the number of seeds $k$ by at least a factor $\ln |\Sigma|$, approximating the objective $\rho$ to within a factor $O(1 - n^{1-\epsilon})$ is $\mathsf{NP}$-hard for all $\epsilon > 0$.

However, when the algorithm does get to exceed the seed set target $k$ by a factor of $\ln |\Sigma|$, much
better bicriteria approximation guarantees can be obtained. 
We recall that a bicriteria algorithm gets to pick more nodes than the optimal solution, but is only judged against the optimum solution with the original bound $k$ on the number of nodes.

Specifically, it is shown that a modification of an algorithm of~\cite{leskovec2007cost} uses $O(k \ln |\Sigma|)$ seeds and finds a seed set whose influence is within a factor $O(1 - n^{1-\epsilon})$ of optimal.
Moreover, two heuristics are investigated.

\begin{itemize}
	\item Run a greedy algorithm to optimize $\rho$ directly, picking one node at a time.
	\item For each objective function $\sigma \in \Sigma$, find a set $S_{\sigma}$ (approximately) 	maximizing $\sigma(S_{\sigma})$. Evaluate each of these sets under $\rho(S_{\sigma})$, and keep the best one.
\end{itemize}

From an experimental perspective, they exhibit instances on which both of the heuristics perform very poorly. 
Then, they focus on more realistic instances , exemplifying the types of scenarios under which robust optimization becomes necessary. 
The main outcome of the experiments is that while the algorithm with robustness as a design goal typically (though not even always) outperforms the heuristics, the margin is often quite small. 
Hence, heuristics may be viable in practice, when the influence functions are reasonably similar.

\section{Deploying IMP on the Field}\label{sec:application}
Social influence maximization has been successfully deployed on the field in different ways~\cite{valente2012network}, e.g., improving nutrition~\cite{kim2015social}, reducing smoking~\cite{starkey2009identifying} and reducing HIV spread~\cite{rice2010positive,yadav2017influence}. Due to the significant increase of importance of the latter issue, in this section we will focus on such an application.

\subsection{Background}
Homeless youth are twenty times more likely to be HIV positive than stably housed youth~\cite{council2012hiv}. 
To reduce rates of HIV infection among youth, many homeless youth service providers conduct peer-leader based social network interventions~\cite{rice2010positive}, where a selected group of homeless youth are trained as peer leaders. This peer-led approach is desirable because service providers have limited resources and homeless youth tend to distrust adults.
The training program of these peer leaders includes detailed information about how HIV spreads and what one can do to prevent infection. They are also taught effective ways of communicating
this information to their peers~\cite{rice2012mobilizing}. 
Because of the limited financial and human resources, service providers can only train a small number of these youth and not the entire population.
As a result, the selected peer leaders in these intervention trainings are tasked with spreading messages about HIV prevention to their peers in their social circles. 
Using these interventions, service providers aim to leverage social network effects to spread information about HIV, and induce behavior change among more and more people in the social network of homeless youth.
Unfortunately, there are further constraints that service providers face, so that they can only train 3-4 peer leaders in every intervention. 
This leads us to do sequential training, where groups of 3-4 homeless youth are called one after another for training, where they are also asked about friendships that they observe in the real-world social network. This information is then used to improve the selection of the peer leaders for the next intervention. 
As a result, the peer leaders for these limited interventions need to be chosen strategically so that awareness spread about HIV is maximized in the social network of homeless youth.

\subsection{Dynamic Influence Maximization under Uncertainty}
We now present two of the most interesting approaches developed to tackle the problem of IMP for HIV prevention. They consider possible uncertainty w.r.t. the existence of the edges, i.e., each edge has an associated probability of existing or not. If an edge exists, then it will have the usual weight indicating the strength of the relationship between the connected individuals.

Let $\mathcal{A} = \{A \subset V s.t. |A| = k\}$ be the set of possible actions that the agent can recommend at every time step $t \in [1,T]$. 
For all $t \in [1,T]$, let $A_t \in \mathcal{A}$ denote the agent’s chosen action in the $t$-th time step.
Upon taking action $A_t$, the agent observes uncertain edges adjacent to nodes in $A_t$, which updates its understanding of the network. 
Moreover, let $G_t$ denote the uncertain network resulting from $G_{t-1}$ with observed (additional edge) information from $A_t$. 
For all $t \in [1,T]$, we define a history $H_t$ of length $t$ as a tuple of past choices and observations: $H_t = \langle G_0, A_1, \ldots, A_{t-1}, G_t \rangle$.
Denote by $\mathcal{H}_t = \{H_k | k \leq t\}$ the set of all possible histories of length less than or equal to $t$. 
Finally, we define an $t$-step policy $\Pi_t: \mathcal{H}_t \rightarrow \mathcal{A}$ as a function that takes in histories of length less than or equal to $t$ and outputs a $k$ node choice for the current
time step. 
As per the diffusion model, they adopt a variant of the IC model, so that nodes get multiple chances to influence their uninfluenced neighbors.

We are now ready to provide the formal definition of the Dynamic Influence Maximization under Uncertainty (DIME)~\cite{yadav2016using}.

\begin{definition}[DIME problem]
	Consider as input an uncertain network $G_0 = (V,E)$ and integers $T$ for the number of rounds and $k$ for the cardinality of the possible leaders that may be selected. 
	Denote by $\mathcal{R}(H_T,A_T)$ the expected total number of influenced nodes at 	the end of stage $T$, given the $T$-length history of previous observations and actions $H_T$, along with $A_T$, the action chosen at time $T$. 
	Let $\mathbb{E}_{H_T,A_T\sim\Pi_T}[\mathcal{R}(H_T,A_T)]$ denote the expectation over the 	random variables $H_T = \langle G_0, A_1, \ldots, A_{T-1}, G_T \rangle$ and $A_T$, where $A_t$ are chosen according to $\Pi_T(H_t)$, for every $t \in [1,T]$, and $G_t$ are drawn according to the distribution over uncertain edges of $G_{t-1}$ that are revealed by $A_t$. 
	The goal of DIME is to find an optimal $T$-step policy $\Pi^*_T = \argmax_{\Pi_T}\mathbb{E}_{H_T,A_T\sim\Pi_T}[\mathcal{R}(H_T,A_T)]$.
\end{definition}

\paragraph{HEALER}
Hierarchical Ensembling based Agent which pLans for Effective Reduction in HIV Spread~\cite{yadav2016using} is a software agent that casts the DIME problem as a Partially Observable Markov Decision Process (POMDP)~\cite{puterman2014markov} to compute a T-step online policy for selecting $k$ nodes for $T$ stages. 
POMDPs are a valuable approach for three reasons. First, service providers select $T$ different subsets of nodes sequentially. Each subset of $k$ nodes is mapped to a unique POMDP action.
Then, the service providers do not see the exact network state at any given point in time. 
HEALER maps each POMDP state to indicate which node is already influenced and which node is not. 
Finally, the observation received by service providers about edges connected to peer leaders is analogous to the observations received in POMDPs. 
HEALER utilizes hierarchical ensembling techniques, creating ensembles of smaller POMDPs at two different levels.
First, the original POMDP is divided into several smaller intermediate POMDPs using graph partitioning techniques. 
Next, each intermediate POMDP is further subdivided into several smaller sampled POMDPs using graph in parallel using novel online planning methods, so that each sampled POMDP executes a Monte Carlo tree search~\cite{silver2010monte} to select the best action in that sampled POMDP. 
The solutions of these sampled POMDPs are combined to form the solution of the intermediate POMDPs. Similarly, the solutions of the intermediate POMDPs are combined to form the solution of the original POMDP.

HEALER consists of two major components: a network generation application for gathering information about social networks and a DIME Solver. 

The Network Generation Application gathers information about social ties in the homeless youth social network by interacting with youth via a network generation application. Once a fixed number of homeless youth register in the network application, HEALER parses the contact lists of all the registered homeless youth on social media and generates the social network between these youth. HEALER adds a link between two people if and only if both people are contacts on social media, and are registered in its network generation application. Unfortunately, there is uncertainty in the generated network as friendship links between people who are only friends in real-life are not captured by HEALER.

The DIME Solver takes as input the approximate social network previously generated, and solves the DIME problem using HEALER algorithm, which provides the solution of this DIME problem as a series of recommendations to homeless shelter officials.

\paragraph{DOSIM}
The Double Oracle for Social Influence Maximization~\cite{wilder2017uncharted} is a novel algorithm that solves a generalization of the DIME problem. 
The key motivation behind DOSIM is to be able to select actions, i.e., set of $k$ nodes, for $T$ stages without knowing the exact model parameters. HEALER dealt with this issue by assuming a specific value based on suggestions by service providers.
DOSIM works with interval uncertainty over the parameters related to the edges, i.e., the existence probability and the weight of the edge. 
This generalizes the model used by HEALER to include higher-order uncertainty over the probabilities in addition to the uncertainty induced by the probabilities themselves. 
DOSIM chooses an action which is robust to this interval uncertainty. Specifically, it finds a policy that achieves close to optimal value regardless of where the unknown probabilities lie within the interval. 
The problem is formalized as zero sum game between the algorithm, which picks a policy, and a \textit{nature-type} adversary who chooses the model parameters. This game formulation represents a key advance over HEALER’s POMDP policy since it enables DOSIM to output mixed strategies over POMDP policies, making it robust against worst-case propagation probability values. 
Moreover, DOSIM receives periodic observations that are used to update its understanding of its belief state, i.e., probability distribution over different model parameters. The strategy space for the game is intractably large because there are an exponential number of policies. 
Thus, DOSIM uses a double oracle approach: by iteratively computing best responses for each player, DOSIM finds an approximate equilibrium of the game without having to enumerate the entire set of policies.

\paragraph{Deployment Process}
We now describe the approach that has been followed to design and test the HEALER and DOSIM algorithms in a real case study.

During the recruitment, the youth take a 20 minute baseline survey, which allows to determine their current risk-taking behaviors (e.g., they are asked about the last time they got an HIV test, etc.). 
After recruitment, the friendship based social network that connects these homeless youth is generated. The authors rely on two information sources to generate this network: online contacts of homeless youth and field observations made by the authors and service providers. Online contacts of homeless youth are used to build a first approximation of the real-world social network of homeless youth.
Then, this network is refined using field observations made by the authors and the service providers. All edges inferred in this manner are assumed to be certain edges.

Next, the generated network is used by the software agents to select actions for $T$ stages. In each stage, an action is selected using the pilot’s intervention strategy. The $k$ peer leaders of this chosen action are then informed about HIV by pilot study staff during the intervention. These peer leaders also reveal more information about newer friendships, which are incorporated into the net work so that the agents can select better actions in the next stage of interventions. 
The follow up phase consists of meetings, where the peer leaders are asked about any difficulties they faced in talking to their friends about HIV. 

Then, during in-person surveys, they are asked if some youth from within the pilot study talked to them about HIV prevention methods, after the pilot study began. Their answer helps determine if information about HIV reached them in the social network or not. Thus, these surveys are used to find out the number of youth who got informed about HIV as a result of our interventions. They are also asked to take the same survey about HIV risk that they took during recruitment. 
These post-intervention surveys allow to compare HEALER, DOSIM and Degree Centrality (DC)~\cite{wasserman1994social} in terms of information spread, i.e., how successful were the agents in spreading HIV information through the social network, and behavior change, i.e., how successful were the agents in causing homeless youth to test for HIV. 
The experiments showed that HEALER and DOSIM’s strategies were able to improve over DC’s information spread by over $184\%$.
Moreover, peer leaders chosen by HEALER and DOSIM converted $37\%$ and $25\%$, respectively, of the youth to HIV testers, while he ones chosen by DC did not convert any youth to testers.

\section{Conclusions and Future Research}\label{sec:future}
In this survey we presented the main results achieved for the influence maximization problem.
We started by introducing the main elements characterizing a social network and then formally defined the influence maximization problem, i.e., the problem of finding the smallest number of individuals in the network able to influence the greatest number of them.

We discussed the main contributions that have been developed w.r.t. the principal features characterizing the problem, namely, the diffusion model, the way of placing seeds from which the influence diffuses, and uncertainties that may affect the structure of the network.
Then, for each case, we reported the most important theoretical results and algorithmic solutions that have been designed, showing how they evolved and have been adapted to various scenarios.
Finally, we presented a real case in which such techniques have been exploited to maximize HIV awareness in youth by identifying natural leaders in different communities. The proposed algorithms have a remarkable behavior w.r.t. the baseline on synthetic data, and showed their true effectiveness also for the pilots that have been conducted.

Next works in this field could follow different directions.
Beside further improving the scalability of the algorithms, novel additional features could be integrated either in the diffusion models or in the ways seeds are placed or accessed in the network. 
Independently by the direction that is explored, we believe that future research lines will be strongly driven by applications since, now more than ever, social networks could raise people's attention to crucial issues, and truly informing people, hoping that in the future we will employ such tools more and more often for social good.

\bibliographystyle{plain}
\bibliography{contents/soc_inf_citations}

\newpage
\appendix
\section{Notation Table} \label{appendix:notation}
We report in~\Cref{tab:notation} the main symbols used throughout the paper.

\begin{table}[!h]
	\centering
	\begin{tabular}{|l|l|l|}
		\cline{2-3}
		\multicolumn{1}{ c| }{} & Symbol & Meaning \\
		\cline{2-3} \hline
		\multirow{19}{*}{\rotatebox{90}{Model}}
		& $G$ & Graph \\ \cline{2-3}
		& $V$ & Set of vertices \\ \cline{2-3}
		& $n$ & Cardinality of $V$ \\ \cline{2-3}
		& $v$ & Vertex \\ \cline{2-3}
		& $E$ & Set of edges \\ \cline{2-3}
		& $m$ & Cardinality of $E$ \\ \cline{2-3}
		& $e$/$(v, v')$ & Edge \\ \cline{2-3}
		& $p(e)$/$p(v,v')$ & Influence probability of edge $e$/$(v,v')$ \\ \cline{2-3}
		& $S$ & Set of seeds \\ \cline{2-3}
		& $s$ & Seed \\ \cline{2-3}
		& $k$ & Cardinality of the initial set of seeds \\ \cline{2-3}    
		& $S^*$ & Optimal set of seeds of cardinality $k$\\ \cline{2-3}
		& $\sigma(\cdot)$ & Influence function \\ \cline{2-3}
		& $\sigma(S)$ & Expected number of nodes influenced by seeds $s \in S$\\ \cline{2-3}
		& $t$ & Time instant\\ \cline{2-3}
		& $\theta_v$ & Activation threshold of node $v$ \\ \cline{2-3}
		& $d(v)$ & Degree of vertex $v$ \\ \cline{2-3}
		& $d^*$ & Highest degree of the vertices in the network \\ \cline{2-3}
		& $CC(v)$ & Clustering coefficient of vertex $v$ \\ 

		\hline
		
		\multirow{4}{*}{\rotatebox{90}{Acronyms}}
		& IMP & Influence maxmization problem \\ \cline{2-3}
		& RIMP & Robust Influence maxmization problem \\ \cline{2-3}
		& IC & Independent Cascade diffusion model \\ \cline{2-3}
		& LT & Linear threshold diffusion model\\
		
		\hline
	\end{tabular}
	\caption[Symbols' table]{Symbols' table.}
	\label{tab:notation}
\end{table}
\end{document}